\documentclass[prd,preprint,superscriptaddress,preprintnumbers,eqsecnum,showpacs,nofootinbib,nobibnotes]{revtex4}
\usepackage{amsfonts,amsmath,amssymb,bm,natbib}
\usepackage{graphicx} 

\newcommand{\be}{\begin{equation}}
\newcommand{\bea}{\begin{eqnarray}}
\newcommand{\ee}{\end{equation}}
\newcommand{\eea}{\end{eqnarray}}

\def\s#1{{\scriptscriptstyle #1}}

\def\noeq#1{(\ref{#1})}
\def\1eq#1{Eq.~(\ref{#1})}

\def\2eqs#1#2{Eqs.~(\ref{#1}) and~(\ref{#2})}
\def\3eqs#1#2#3{Eqs.~(\ref{#1}),~(\ref{#2}) and~(\ref{#3})}

\def\fig#1{Fig.~\ref{#1}}

\def\DeltaQ{\Delta_\s{\nf}}
\def\mQ{m^2_\s{\nf}}

\def\diff#1{{\rm d}^#1}

\def\ie{{\it i.e.}, }
\def\eg{{\it e.g.}, }

\def\n#1{({\it #1}\,)}

\def\nf{N_{\!f}}

\def\Y{Y}

\def\quark{\widehat{X}}

\def\pslash{p\hspace{-0.18cm}\slash} 
\def\qm{{\cal M}}

\begin{document}

\title{Gluon mass generation in the presence of dynamical quarks}

\author{A.~C. Aguilar}
\affiliation{University of Campinas - UNICAMP, 
Institute of Physics ``Gleb Wataghin'' \\
13083-859 Campinas, SP, Brazil}

\author{D. Binosi}
\affiliation{European Centre for Theoretical Studies in Nuclear
Physics and Related Areas (ECT*) and Fondazione Bruno Kessler, \\Villa Tambosi, Strada delle
Tabarelle 286, 
I-38123 Villazzano (TN)  Italy}

\author{J. Papavassiliou}
\affiliation{\mbox{Department of Theoretical Physics and IFIC, 
University of Valencia and CSIC},
E-46100, Valencia, Spain}

\begin{abstract}
We study  in detail the impact  of dynamical quarks on  the gluon mass
generation mechanism,  in the  Landau gauge, for  the case of  a small
number  of quark families.   
As in  earlier considerations,  we assume
that  the  main bulk  of  the  unquenching  corrections to  the  gluon
propagator originates from the  fully dressed quark-loop diagram.  The
nonperturbative evaluation  of this diagram provides  the key relation
that expresses the unquenched gluon propagator as a deviation from its
quenched  counterpart.  This  relation is  subsequently coupled  to the
integral  equation  that  controls   the  momentum  evolution  of  the
effective gluon  mass, which  contains a single  adjustable parameter; 
this constitutes a major improvement compared to the analysis presented  
in Phys.~Rev. D86 (2012) 014032, where the 
behaviour of the gluon propagator in the deep infrared was estimated through numerical 
extrapolation. 
The resulting  nonlinear system is then  treated numerically, yielding 
unique solutions  for the modified  gluon mass and the  quenched gluon
propagator, which fully confirm the picture put forth recently 
in several continuum  and  lattice  studies. In particular, 
an infrared  finite  gluon  propagator  emerges, whose  
saturation  point is considerably  suppressed, due to a 
corresponding increase in the value of the gluon  mass.   
This characteristic feature becomes more pronounced as the number of active
quark  families increases,  and  can be deduced from the  infrared
structure of the kernel entering in the gluon mass equation.

\end{abstract}


\maketitle

\section{Introduction}

The concept of a momentum-dependent gluon mass~\cite{Cornwall:1981zr}
has been the focal point of considerable attention in recent years, mainly because it offers 
a natural and self-consistent explanation~\cite{Aguilar:2008xm} 
for the infrared finiteness of the (Landau gauge) gluon propagator and ghost 
dressing function, observed in large-volume lattice simulations, both in  
$SU(2)$~\cite{Cucchieri:2007md} and in $SU(3)$~\cite{Bogolubsky:2009dc}. 
Given the purely nonperturbative nature of the associated mass generation mechanism, the 
Schwinger-Dyson equations (SDEs) constitute the most appropriate framework for 
studying such phenomenon in the continuum~\cite{Binosi:2007pi,Alkofer:2000wg,Fischer:2006ub,Braun:2007bx}. 

At the level of the SDEs, the general analysis 
finally boils down to the study of an 
integral equation, 
to be referred as the \textit{mass equation}~\cite{Aguilar:2011ux,Binosi:2012sj}, which governs the evolution 
of the dynamical gluon mass, $m^2(q^2)$, as a function of the momentum $q^2$ (see  \1eq{masseq} below).
It turns out that this special equation 
has been derived {\it exactly}, 
following the elaborate procedure explained in detail in~\cite{Binosi:2012sj}. 
This particular construction was carried out 
within the framework provided by the synthesis of the 
pinch technique (PT)~\cite{Cornwall:1981zr,Cornwall:1989gv,Pilaftsis:1996fh, 
Binosi:2002ft,Binosi:2003rr,Binosi:2009qm} with the background field method (BFM)~\cite{Abbott:1980hw}, known in the 
literature as the PT-BFM scheme~\cite{Aguilar:2006gr,Binosi:2007pi,Binosi:2008qk}.

This exact mass equation, however, may not be treated in its complete form, because a particular field-theoretic ingredient entering in it 
is not fully known. Therefore, an approximate 
version of the full equation has been considered instead~\cite{Binosi:2012sj}, 
which depends on a free adjustable parameter. 
The detailed numerical study of this particular equation, 
for pure $SU(3)$ Yang-Mills, revealed that its   
solutions depend strongly on the  
precise shape of the gluon propagators entering in its kernel.  
In particular, the $q^2$ region below 1 GeV$^2$ appears to 
be crucial for obtaining physically acceptable 
(\ie positive-definite and monotonically decreasing) solutions.

Interestingly enough, as has been shown in recent SDEs~\cite{Aguilar:2012rz} and lattice  studies~\cite{Ayala:2012pb}, 
the gluon propagator, $\Delta(q^2)$, undergoes considerable changes in this particular 
region of momenta 
when a small number of light (dynamical) quarks is included in the theory, \ie when  
``unquenching'' takes place.
This basic observation, in turn, motivates the further scrutiny of the mass equation 
under the special conditions that occur    
when the transition from pure Yang-Mills to real world QCD takes place. 
In particular, it is clearly of the outmost importance 
to establish, in quantitative detail, how the inclusion of dynamical quarks 
affects the entire mechanism of gluon mass generation.  
In fact, from the point of view of computing unquenching effects to the gluon propagator, 
the present work constitutes a sequel to~\cite{Aguilar:2012rz}, 
whose main improvement is the incorporation of the gluon mass equation into the general 
dynamical analysis 
(in the context of the so-called ``scaling'' solutions, where no mass scale is generated, 
the unquenching effects  
have been investigated in detail in~\cite{Fischer:2003rp,Fischer:2005en}). 
 
If one subscribes to the interpretation that the  
infrared finiteness of the lattice propagators
is a consequence of the generation of such a mass, 
then the new unquenched lattice results of~\cite{Ayala:2012pb} would seem to indicate that 
the general mass generation picture persists 
in the presence of quark loops. In fact, the observed considerable 
suppression of the value of the 
saturation point of the unquenched propagators with respect to the 
quenched ones, would clearly suggest that 
the corresponding gluon mass increases [since $\Delta^{-1}(0)= m^2(0)]$. 
To be sure, this basic argument must be supplemented by a more careful 
analysis, taking into account the effects of renormalization. In particular, 
the required renormalization of the gluon propagators by a multiplicative constant 
may shift the location of their corresponding saturation points. 
However, provided that all curves are renormalized at the same (subtraction) point,  
the relative size (ratio) between these saturation points will remain unaltered, 
indicating the aforementioned relative increase of the effective gluon mass.

To explore this fundamental issue systematically, 
one needs a faithful description of the 
effects of the inclusion of quark-loops to the gluon propagator.
To be sure, one could envisage the possibility of substituting the 
unquenched lattice data directly into the mass equation, in a spirit similar to that 
followed in the pure Yang-Mills case. 
However, we will refrain from using this latter set of data, and will resort instead to the  full
SDE approach presented in~\cite{Aguilar:2012rz}, which allows the 
approximate inclusion of such effects in the quenched propagator.

This particular approach assumes the presence of a small number of quark families, 
and treats the entire effect of ``unquenching'' as a ``perturbation'' to the quenched gluon propagator\footnote{In the case of several quark families the changes induced may be rather 
profound, see,\eg~\cite{DelDebbio:2010zz,Cheng:2011qc}, and references therein.}.
Then, the main bulk of the ``unquenching'' is attributed to the {\it fully dressed} quark-loop graph, 
while higher loop contributions are considered to be subleading.
The inclusion of this particular nonperturbative loop induces considerable 
modifications to the shape of the resulting propagator, but does 
not provide precise information on the existence and value 
of the new saturation point. Indeed, such information may be only gathered in conjunction with the 
gluon mass equation, which was unavailable at the time of the original analysis.
As a result, in the work of~\cite{Aguilar:2012rz} 
the existence of a saturation point was {\it assumed}, and its approximate value  was obtained through 
simple extrapolation from the overall shape of the propagator in the region around 0.05 GeV$^2$.

Here, instead, our present knowledge of the mass equation allows us to  
carry out a thorough analysis of this particular dynamical question.
Specifically, the equation describing  the unquenching of the gluon propagator is coupled to the mass equation, and the resulting non-linear system is solved numerically.
In this way, one eventually obtains as a result 
the unquenched gluon propagator in the entire range of physically relevant momenta, 
including the value of its saturation point.

The main result of this article may be summarized as follows.
The solution of the system reveals that the mass equation yields 
physical solutions in the presence of quarks, specifically  
for the typical cases $\nf=2$ (two degenerate light quarks), 
and $\nf=2+1+1$ (two degenerate light quarks and two heavy ones),  considered in 
the recent lattice simulations~\cite{Ayala:2012pb}. 
In fact, one establishes clearly that 
the value of the gluon    
mass at the origin increases as additional quarks are included in the theory.
In addition, one obtains a concrete prediction for the quenched gluon propagator,  which compares 
rather favorably with the aforementioned lattice results. We emphasize that 
throughout this analysis we make use of a single adjustable parameter, 
which is meant to model contributions to the mass equations stemming from the fully-dressed three gluon vertex. 

The article is organized  as follows. In Section~\ref{unq-mess}, 
we first define the  basic quantities  appearing in the problem, and 
introduce  some  important relations  characteristic  to the  PT-BFM
framework. Then, we 
describe the modifications that must  
be implemented at the level of the gluon SDE  in order to consistently accommodate
a dynamical gluon mass. Finally,  we 
review the theoretical origin and properties of the two key dynamical equations,  
namely the gluon mass (integral) equation and the 
unquenching formula. 
Section~\ref{numan} is dedicated to the solution of the 
system formed when the aforementioned two equations are coupled to each other.
To that end,  
we begin by reviewing the  main ingredients
appearing in them, and spell out the approximations 
employed in their treatment. 
Then we 
proceed to  the numerical solution,
describing in  detail the iterative procedure  adopted.  
The numerical
results obtained are then compared with the latest
lattice  results, and some of the possible reasons for the 
observed deviations are discussed. 
Our   conclusions    are finally   presented   in
Section~\ref{conc}.

\section{\label{unq-mess}Gluon Mass equation with unquenched gluon propagators}

The full gluon propagator (quenched or unquenched) in the Landau gauge assumes the general form 
\be
i{\Delta}_{\mu\nu}(q)=-i{\Delta}(q^2)P_{\mu\nu}(q); \qquad 
P_{\mu\nu}(q)=g_{\mu\nu}-{q_\mu q_\nu}/{q^2},
\label{prop-def}
\ee 
where $\Delta(q^2)$ is related to the scalar form factor of the  
gluon self-energy $\Pi_{\mu\nu}(q)=\Pi(q^2)P_{\mu\nu}(q)$  through
\be
\Delta^{-1}(q^2)=q^2+i\Pi(q^2) \,.
\ee 
In addition, the corresponding \textit{inverse} gluon dressing function, $J(q^2)$, is defined as
\begin{equation}\label{gldressing}
\Delta^{-1}(q^2)=q^2 J(q^2), 
\end{equation}
and, as has been explained in~\cite{Aguilar:2009ke,Aguilar:2011ux},  it is intimately related with the 
definition of the renormalization-group invariant QCD effective charge.  

As has been demonstrated in detail in the recent literature~\cite{Aguilar:2011xe,Ibanez:2012zk}, 
the nonperturbative dynamics of Yang-Mills theories gives rise to a dynamical (momentum-dependent) 
gluon mass, which accounts for the infrared finiteness of $\Delta(q^2)$. Specifically, 
from the kinematic point of view, we will describe the transition 
from a massless to a massive gluon propagator by carrying out the replacement 
$\Delta^{-1}(q^2) \to \Delta_m^{-1}(q^2)$, where (Minkowski space) 
\begin{equation}
\label{massiveprop}
\Delta_m^{-1}(q^2) = q^2 J_m(q^2) - m^2(q^2).
\end{equation}
Notice that the subscript ``$m$'' indicates that effectively one has now a mass inside the corresponding 
expressions: for example, whereas perturbatively $J(q^2)\sim \ln q^2$, after dynamical gluon mass 
generation has taken place, one has $J_m(q^2)\sim \ln(q^2 + m^2)$.
The presence of this mass, in turn, tames the Landau pole appearing in the (perturbative)  
effective charge, causing its saturation in the deep infrared~\cite{Aguilar:2009nf,Aguilar:2010gm}. 

From the dynamical point of view, 
the emergence of massive solutions from the gluon propagator SDE depends crucially on  
the inclusion of a set of special 
vertices, to be generically denoted by $V$ and called \textit{pole vertices}~\cite{Aguilar:2011ux},
which contain massless, longitudinally coupled poles.
 These vertices are added to the usual (fully dressed) 
vertices of the theory, and have a two-fold effect:
they trigger the well-known Schwinger mechanism, thus endowing the gluon with a dynamical mass,   
while, at the same time,  
they guarantee that the 
Slavnov-Taylor identities 
(STIs) of the theory maintain exactly the same form before and after the mass has been dynamically generated. 
In particular, the transversality of $\Pi_{\mu\nu}(q)$ 
in the presence of masses is preserved, provided that  
one replaces all vertices $\Gamma$  appearing in the gluon SDE by 
\be
\Gamma \longrightarrow \Gamma' = \Gamma_m + V,
\label{replacever}
\ee
with $V$ having the precise structure that will make the 
new vertices $\Gamma'$ satisfy the same formal STIs as the $\Gamma$ before, but with the replacement
$\Delta^{-1} \longrightarrow \Delta_m^{-1}$. 

It turns out that 
$J_m(q^2)$ and $m^2(q^2)$ satisfy  two separate, but \textit{coupled}, 
integral equations of the generic type\footnote{Note that in the rest of this article we will suppress the subscript ``$m$'' in $\Delta$; 
it is understood that the gluon propagator is infrared finite.}~\cite{Aguilar:2011ux}
\begin{eqnarray}
&& J_m(q^2) = 1 + \int_k {\cal K}_1(k,q,m^2,\Delta), 
\nonumber \\
&& m^2(q^2) = \int_k {\cal K}_2(k,q,m^2,\Delta), 
\label{meq}
\end{eqnarray}
such that ${\cal K}_1, {\cal K}_2\neq 0$, as $q\rightarrow 0$. In the  equations above we have introduced the dimensional regularization measure $\int_k=\mu^\epsilon\int\!\diff{d}k/(2\pi)^d$ where $d=4-\epsilon$ is the space-time dimension and $\mu$ the 't Hooft mass.


\begin{figure}[!t]
\includegraphics[scale=.95]{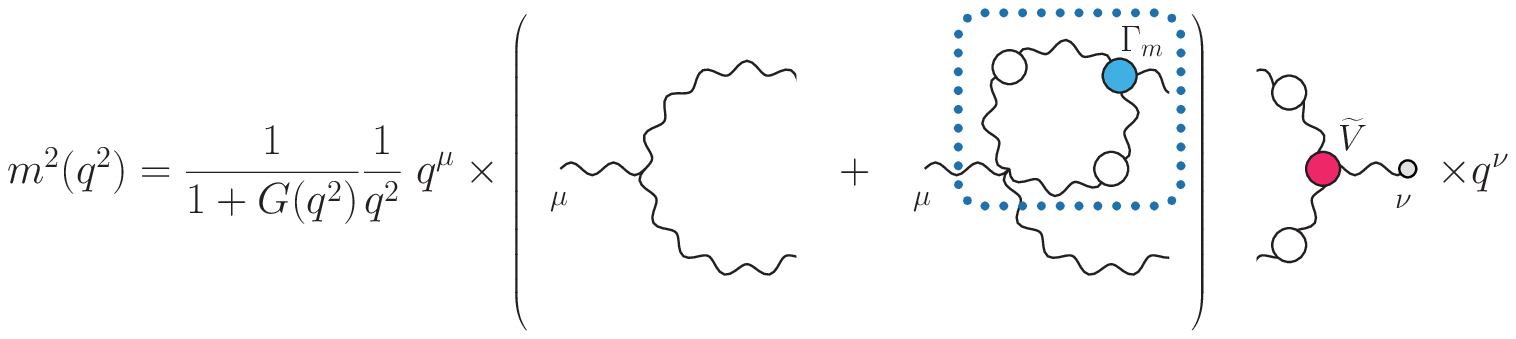}
\caption{\label{meq-fig} The effective SDE satisfied by the dynamical gluon mass. The blue vertex corresponds to the vertex $\Gamma_m$ of \1eq{replacever}, while the red vertex~$\widetilde{V}$ indicates a pole vertex in which one of the gluon is leg is a background leg. Finally,  the box singles out the quantity $Y(k^2)$ which represents the purely two-loop dressed correction to the one-loop dressed mass equation kernel.}
\end{figure}


The exact closed form of the kernel ${\cal K}_2$, and hence the full  
integral equation that determines $m^2(q^2)$, has been derived in~\cite{Binosi:2012sj}, 
by appealing to the special properties of the pole vertices $V$.
The resulting homogeneous equation is given by~\cite{Binosi:2012sj} (see also~\fig{meq-fig}) 
\be
m^2(q^2) = -\frac{g^2 C_A}{1+G(q^2)}\frac{1}{q^2}\int_k m^2(k^2) \Delta_\rho^\mu(k)\Delta^{\nu\rho}(k+q)
{\cal K}_{\mu\nu}(k,q),
\label{masseq}
\ee
with 
\bea
{\cal K}_{\mu\nu}(k,q) &=& [(k+q)^2 - k^2] \left\{ 1 - [\Y(k+q) + \Y(k)]\right\}g_{\mu\nu}
\nonumber \\
&+& [\Y(k+q)-\Y(k)](q^2 g_{\mu\nu}-2q_\mu q_\nu). 
\label{massK}
\eea
The quantity $Y$ represents the subdiagram nested inside the 
two-loop dressed graph of \fig{meq-fig},  given by 
\be
\Y(k^2)=\frac{g^2 C_A}{4 k^2} \,k_\alpha\! \int_\ell\!\Delta^{\alpha\rho}(\ell)
\Delta^{\beta\sigma}(\ell+k)\Gamma_{\sigma\rho\beta}(-\ell-k,\ell,k),
\label{theY}
\ee
with $\Gamma_{\sigma\rho\beta}$  the full three-gluon vertex, 
and $C_A$ the Casimir eigenvalue in the adjoint representation [$C_A=N$ for $SU(N)$].

Finally, the function $G(q^2)$ corresponds to the 
$g_{\mu\nu}$ component of a special two point-function, which constitutes a
crucial ingredient in a set of 
powerful identities, relating the conventional Green's functions to those of the BFM~\cite{Grassi:1999tp,Binosi:2002ez}.  
For the case of the conventional gluon propagator, $\Delta$, and the PT-BFM gluon propagator, denoted by $\widehat\Delta$, 
the relevant identity reads 
\begin{equation}\label{propBQI}
\Delta(q^2) = [1+G(q^2)]^2 \widehat{\Delta}(q^2);
\end{equation}
its application  
at the level of the corresponding SDEs leads to the appearance 
of the factor $1+G(q^2)$ in \1eq{masseq}. 

The numerical treatment of \1eq{masseq}, under certain simplifying assumptions regarding the form of $\Y(k^2)$ (see next section),
gave rise to solutions that display the basic qualitative features expected on general field-theoretic considerations, and 
employed in numerous phenomenological studies. In particular, the $m^2(q^2)$ obtained are monotonically decreasing functions 
of the momentum and vanish rather rapidly in the ultraviolet~\cite{Cornwall:1981zr,Binosi:2012sj}.

As explained in the Introduction, the purpose of the present work is 
to study the effects induced on the gluon mass by 
the inclusion of a small number of light quark families.
To that end,   
one needs to solve \1eq{masseq} using unquenched gluon propagators, $\DeltaQ(q^2)$, namely 
carrying out the substitution $\Delta(q^2) \to \DeltaQ(q^2)$
inside the integral of the r.h.s.. 

\begin{figure}[!t]
\includegraphics[scale=0.7]{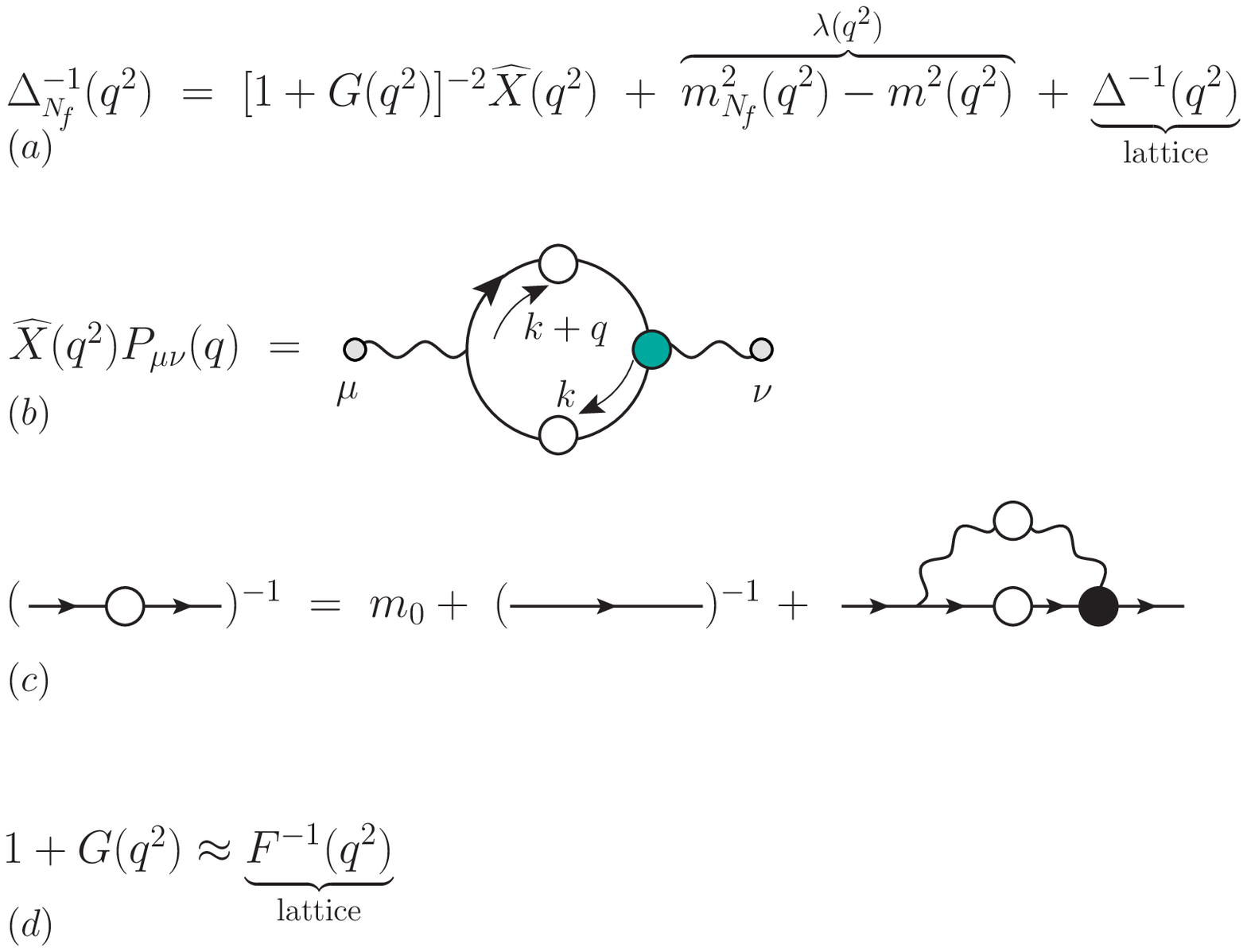}
\caption{\label{Unq-gl-SDE} Schematic representation of the unquenched propagator $(a)$,
corresponding to \1eq{mastformeuc}, and some of the ingredients [$(b)$, $(c)$, and $(d)$] entering in it.
In particular, $(b)$ represents the quark loop, which, in the approximation employed, is the
only source of quark-dependence.
The quark propagators entering in $(b)$ are solutions of the gap equation
depicted in $(c)$, where $m_0$ denotes the appropriate current mass.
Finally, the function $G(q^2)$ is obtained from the relation shown in $(d)$, namely~\1eq{funrelapp}.
The quantities obtained from the lattice are also indicated. Note that 
the term $\lambda^2(q^2)$ will be determined {\it dynamically}, once the
mass equation of Fig.~\ref{meq-fig} is coupled to $(a)$.} 
\end{figure}

The corresponding expression for $\DeltaQ(q^2)$ will be obtained following 
the SDE approach presented in~\cite{Aguilar:2012rz}.
Specifically, one assumes that the main bulk of the unquenching
effect is captured by the (fully dressed) one-loop diagram
of~\fig{Unq-gl-SDE} (b), neglecting, at this level of approximation,   
all contributions stemming from (higher order) diagrams 
containing nested quark loops. 

 
Denoting the contribution of this special diagram by 
\be
\widehat{X}_{\mu\nu}(q)=\widehat{X}(q^2)P_{\mu\nu}(q), 
\label{Drnf}
\ee 
we have that                      
\be
\widehat{X}(q^2)=-\frac{g^2}{6}\!\int_k\mathrm{Tr}\left[\gamma^\mu S(k)\widehat\Gamma_\mu(k,-k-q,q)S(k+q)\right].
\label{quark-loop-full}
\ee 
where $S$ denotes the {\it full} quark  propagator, and 
the vertex $\widehat{\Gamma}_\mu$ corresponds to the PT-BFM quark-gluon vertex, satisfying the QED-like Ward identity
\be
iq^\mu \widehat{\Gamma}_\mu(k,-k-q,q)=S^{-1}(k)-S^{-1}(k+q).
\label{GWI}
\ee
Of course, in the case of including various quark loops, corresponding to different quark flavors $\nf$,  the term $\quark^{\mu\nu}(q)$ in \1eq{Drnf}  
is replaced simply by the sum over all quark loops, \ie 
\be
\quark^{\mu\nu}(q) \to \sum_{f}\widehat{X}_{f}^{\mu\nu}(q).
\label{sumX}
\ee

Then, the detailed analysis of~\cite{Aguilar:2012rz} reveals that 
the unquenched gluon propagator $\DeltaQ(q^2)$ may be expressed as a deviation from the  
quenched propagator $\Delta(q^2)$, namely (Euclidean space)
\be
\DeltaQ(q^2) = \frac{\Delta(q^2)}
{1 + \left\{ \quark(q^2) \left[1+G(q^2)\right]^{-2}+ \lambda^2(q^2)\right\}\Delta(q^2)}, 
\label{mastformeuc}
\ee
where the quantity 
\be
\lambda^2(q^2)=\mQ(q^2)-m^2(q^2),
\label{lambda}
\ee 
measures the difference induced to the gluon mass due to the inclusion of quarks.

An important point, which can be established formally and confirmed numerically~\cite{Aguilar:2012rz},  is that the nonperturbative $\widehat{X}(q^2)$ vanishes at the origin, $\widehat{X}(0)=0$, exactly as it happens in perturbation theory. This implies that the mass equation is not  {\it directly} affected by the presence of dynamical quarks, as the methodology used to derive~\3eqs{masseq}{massK}{theY} is left invariant by the unquenching procedure. However,  the solutions of the corresponding equation will be different from those obtained in the quenched case, as the kernel ${\cal K}_{\mu\nu}$ will change, due to the modifications induced by $\widehat{X}(q^2)$ in the overall shape of the propagator throughout the entire range of momenta. Thus, the inclusion of quark loops affects the value of the saturation point of the gluon propagator, not {\it directly} through the presence of the $\widehat{X}(q^2)$, rather {\it indirectly} through the generation of a non vanishing mass difference $\lambda^2(q^2)$.

The next nontrivial step is therefore to treat the mass equation \1eq{masseq} and the unquenching master formula of \1eq{mastformeuc} as a coupled system, and determine both $\mQ$ and $\DeltaQ(q^2)$. This rather involved task will be undertaken in detail in the next section.

\section{\label{numan}Numerical analysis}

Before entering into the technical details of the solution of the system, 
it is important to comment on a fundamental qualitative difference between the present situation and the 
treatment followed in~\cite{Binosi:2012sj}.
 Specifically, as already explained in the previous section, even in the absence of quarks, \1eq{masseq} 
forms part of a system of coupled equations, namely that of \1eq{meq}.  
However, given that the 
corresponding equation for $J_m (q^2)$ is only partially known, the approach adopted in~\cite{Binosi:2012sj}  
was to treat the gluon propagators  
appearing in \1eq{meq} as external quantities, using the lattice results of~\cite{Bogolubsky:2009dc} for their functional form. 
As a consequence, the intrinsically non-linear 
equation~\noeq{meq} (recall~\1eq{massiveprop}) was converted into a linear one. 
As a result of this linearity, one obtained a continuous family of solutions, parametrized by the values of 
a multiplicative constant; the actual solution 
chosen was the one that 
reproduced the saturation point of the lattice  gluon  propagator that was
used as input in \1eq{meq}.  

Here, instead, even though we still decouple the equation for $J_m (q^2)$,  the mass equation retains its 
non-linear nature due to the form of the unquenched gluon propagator, $\DeltaQ(q^2)$, that will be inserted in it. 
Specifically, as can be appreciated from \1eq{mastformeuc}, the $\lambda^2(q^2)$ appearing in the 
denominator of $\DeltaQ(q^2)$  
introduces a non-linear term into \1eq{meq}. Consequently, the solution obtained will be unique, 
and the saturation point of the resulting $\DeltaQ(q^2)$ will emerge as a prediction of the theory rather than as an input from the lattice.    

\subsection{Main ingredients and basic assumptions}

We next proceed to a brief description of the ingredients entering into the two equations~\noeq{meq} and~\noeq{mastformeuc} composing the system under study, together with an account of the most important underlying assumptions. 

\n{i} The determination of the quantity $\widehat{X}(q^2)$ is of central importance, since  
it approximates the entire effect of the inclusion of the active quarks. In order to evaluate it from 
\1eq{quark-loop-full}, one needs specific expressions for the full quark propagator $S$
and the fully-dressed vertex $\widehat\Gamma$. Evidently, the total $\widehat{X}(q^2)$ emerges as the 
sum of the individual flavor contributions, in accordance with \1eq{sumX}.  

The nonperturbative form of each quark propagator entering into $\widehat{X}(q^2)$ 
is obtained from the standard gap equation~\cite{Aguilar:2010cn}, supplemented by  
an appropriate current mass term, in order to make contact with the lattice results of~\cite{Ayala:2012pb}.
In this latter simulation,   
the gluon (and ghost) propagators have been evaluated from large volume configurations (up to $3^3\times6$ [fm$^4$]),  
generated from a lattice action that included (twisted mass) fermions. 
Specifically, one employed two  light degenerate quarks ($\nf=2$),  
with a current mass ranging from 20 to 50 MeV, or two light and two heavy quarks ($\nf=2+1+1$), 
with a strange (charm) quark current mass  roughly set to 95 MeV (1.51 GeV). 
  
\begin{figure}[!t]
\includegraphics[scale=1]{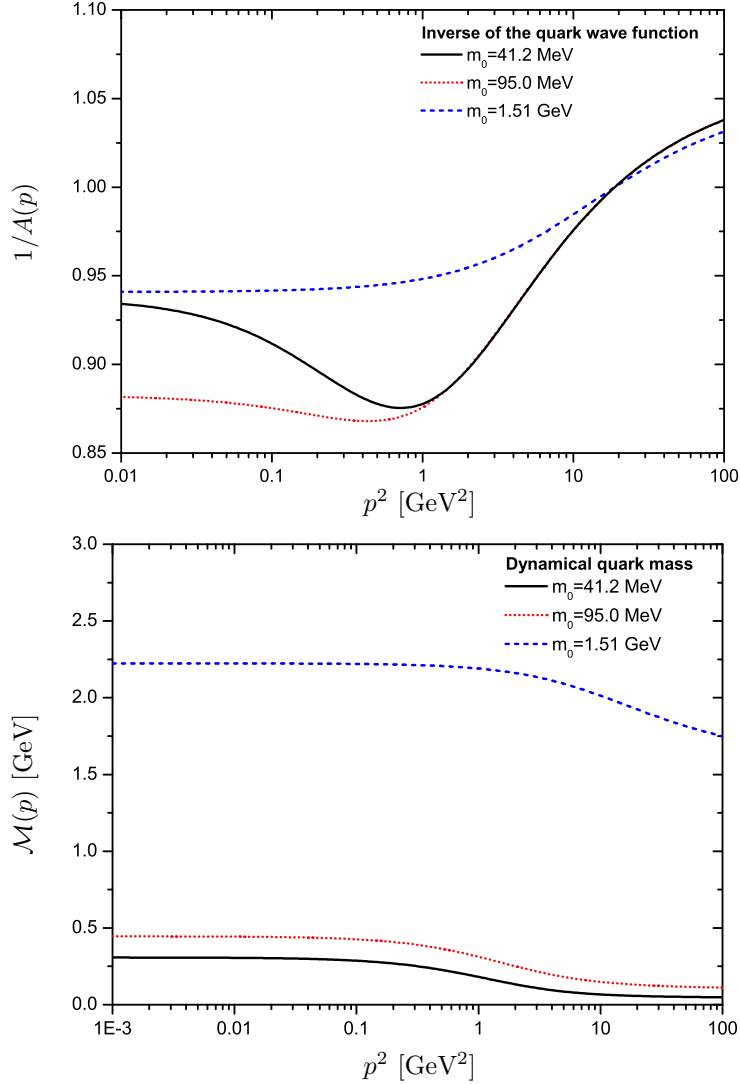}
\caption{\label{CSBquark-mass_wave}(color online)
The inverse of the quark wave-functions (top), and dynamical  quark masses (bottom), obtained from 
the quark gap equation for three different values of the 
current mass: $m_0=41.2$ MeV (black, continuous), $m_0=95$ MeV (red, dotted) and $m_0=1.51$ GeV (blue, dashed). } 
\end{figure}  


For the quark propagator, $S(p)$, we use the standard parametrization 
\be
S^{-1}(p)=-i\left[A(p)\pslash-B(p)\right]=-iA(p)\left[\pslash-\qm(p)\right],
\label{qprop}
\ee
where $A(p)$ represents the ``quark wave-function'', while the ratio $\qm(p)=B(p)/A(p)$ denotes the dynamical quark mass, 
and extracts them  from the corresponding system of integral equations. 
The inverse of the quark wave-functions, $1/A(p)$, and the dynamical quark masses, ${\mathcal M}(p)$, used in the present work, 
are shown in the top and bottom panels of~\fig{CSBquark-mass_wave}, respectively.
They have been obtained from the gap equation of~\cite{Aguilar:2010cn}, where the quenched lattice data 
of~\cite{Bogolubsky:2009dc} were used as input.

As can be seen from~\fig{CSBquark-mass_wave} the two degenerate light quarks acquire a physical mass around $300$ MeV, while the two heavier ones get physical masses around 440 MeV (strange) and 2.2 GeV (charm). It should be stressed that these values for the quark masses are consistent with the ones generally employed in phenomenological calculations~\cite{Maris:2003vk,Wilson:2011aa}.

As for the PT-BFM quark-gluon vertex $\widehat\Gamma$, 
the fact that it satisfies the Abelian Ward identity of \1eq{GWI}  
allows one to  model it by means of the  standard Abelian 
Ans\"atze existing in the literature~\cite{Ball:1980ay,Curtis:1990zs}. In particular, 
$\widehat\Gamma$ is expressed entirely in 
terms of the quantities $A(p)$ and $\qm(p)$, with no reference to auxiliary ghost Green's functions, 
a fact that simplifies considerably the task of computing $\widehat{X}(q^2)$. 

\begin{figure}[!t]
\includegraphics[scale=1]{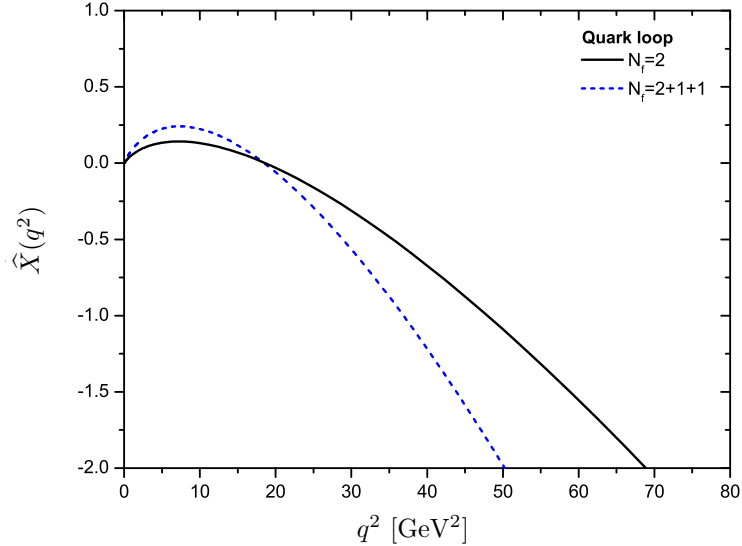}
\caption{\label{quark-loop} The full nonperturbative quark loop contribution $\widehat{X}(q^2)$ for the two cases $\nf=2$ (black, continuous) and $\nf=2+1+1$ (blue, dashed).}
\end{figure}

When all the aforementioned ingredients are put together, the two different  
$\widehat{X}(q^2)$, corresponding to the cases 
$\nf=2$ (black, continuous)
and $\nf=2+1+1$ (blue, dashed), are shown in~\fig{quark-loop}.   
 
Note finally that, as one can establish formally and confirm numerically~\cite{Aguilar:2012rz},  the nonperturbative $\widehat{X}(q^2)$ vanishes at the origin, $\widehat{X}(0)=0$,
exactly as it happens in perturbation theory. 
Evidently, the inclusion of quark loops 
does not affect {\it directly} the value of the saturation point of the gluon propagator. 
However, as already explained, the saturation point will be {\it indirectly} affected, due to 
the modifications induced by $\widehat{X}(q^2)$ in the overall shape of the propagator, throughout the entire range of momenta.


\begin{figure}[!t]
\includegraphics[scale=1]{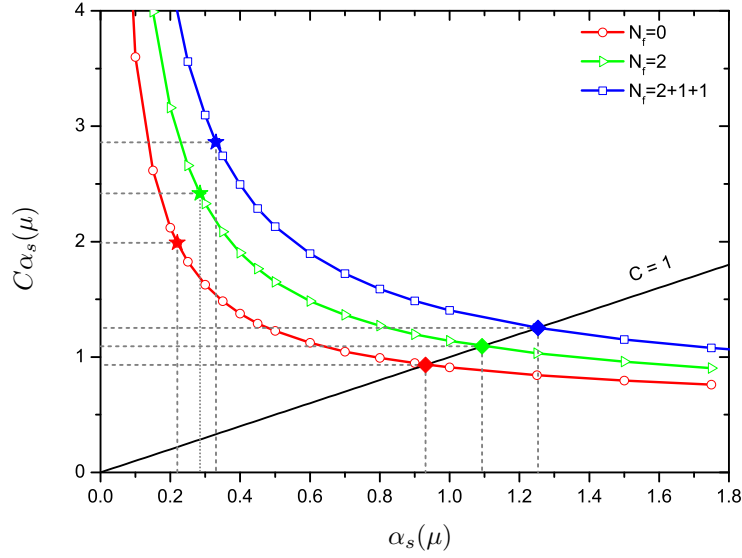}
\caption{\label{Cvsalphaf}  The curves  formed by the  set of  the
pairs  $(C\alpha_s,\alpha_s)$ yielding physical solutions  to the
full  mass equation~\noeq{masseq},  for  the various  numbers of  quark
families $\nf$:  quenched (red, dots), $\nf=2$  (green, triangles) and
$\nf=2+1+1$ (blue,  squared). 
 At the values of $C$ indicated by the stars, 
the solutions obtained corresponds to the expected values of the strong coupling 
$\alpha_s$ at the scale chosen ($\mu=4.3$ GeV); one  has
$C=9.04$   for  $\nf=0$   ($\alpha_s=0.22$),   $C=8.48$  for   $\nf=2$
($\alpha_s=0.285$),       and      $C=8.64$       for      $\nf=2+1+1$
($\alpha_s=0.33$). Finally, the values  of $\alpha_s$ obtained for the
case  in which $Y$  is simply  kept at  the lowest  order perturbative
value  (\ie  $C=1$)  are  $\alpha_s=0.93$  ($\nf=0$),  $\alpha_s=1.09$
($\nf=2$), and $\alpha_s=1.25$ ($\nf=2+1+1$).}
\end{figure} 


\n{ii}  As already mentioned, 
the quenched lattice data for $\Delta(q^2)$~\cite{Bogolubsky:2009dc} 
constitute the starting point for obtaining from \1eq{mastformeuc}
a prediction for $\DeltaQ(q^2)$. In addition, in the Landau gauge,
the quantity $1+G(q^2)$, appearing in both \2eqs{meq}{mastformeuc},
is linked to the inverse of the ghost dressing function $F(q^2)$
through~\cite{Grassi:2004yq,Aguilar:2009pp}
\be
F^{-1}(q^2) \approx 1+G(q^2).
\label{funrelapp}
\ee
This relation, which is valid to a very good approximation, and becomes an exact equality at $q^2=0$,  
allows one to use the lattice results of~\cite{Bogolubsky:2009dc} for the ghost dressing function, in order to determine $G(q^2)$. 
Notice that both sets of data will be renormalized at the 
scale $\mu=4.3$~GeV, which is the last available point in the ultraviolet.   
At this scale, the expected value of the (quenched) strong effective charge is $\alpha_s=g^2/4\pi=0.22$~\cite{Boucaud:2005rm,Boucaud:2008gn}.

\n{iii}
The function $\Y(k^2)$  represents a crucial ingredient 
of \1eq{masseq}. However, its exact closed form is not available, 
mainly  because our present knowledge of the full three-gluon vertex, entering in its definition,     
is relatively limited; we must therefore resort to suitable approximations.
In particular, we will employ the lowest-order perturbative expression for  $\Y(k^2)$,
obtained from \1eq{theY} by substituting the tree-level values for all quantities appearing there.  
Within this approximation, and after carrying out momentum subtraction renormalization (MOM) at $k^2=\mu^2$, one obtains~\cite{Binosi:2012sj}
\be
Y_{\mathrm R}(k^2)=-\frac {\alpha_s C_A}{4\pi}\frac{15}{16}\log\frac{k^2}{\mu^2},
\label{Yappr}
\ee
where $\alpha_s$ is the value of the coupling at the subtraction point chosen.
This simple approximation will be compensated, in part,  
by multiplying $Y_{\mathrm R}(k^2)$ by and arbitrary constant $C$, 
\ie by implementing the replacement 
$Y_{\mathrm R}(k^2) \to C\, Y_{\mathrm  R}(k^2)$
treating $C$ as a free parameter.
In this heuristic way, one hopes to model 
further  corrections that may  be added  to the
``skeleton'' result  provided by~\1eq{Yappr}. This  particular procedure has been proved  
sufficient  for obtaining in the quenched case physically  meaningful  solutions  for $m^2(q^2)$     
for    reasonable     values     of    the     constant $C$~\cite{Binosi:2012sj}. 
As the numerical study presented in~\fig{Cvsalphaf} shows, this continues to be true also in the unquenched case. 

\n{iv} A particularly helpful constraint may be obtained by taking the $q^2\to0$ limit of the mass equation~\noeq{masseq}, 
which gives (passing to spherical coordinates, and setting $y=k^2$)
\be
m^2(0)=-\frac{3}{8\pi}\alpha_s C_A\, F(0)\int_0^\infty\! \mathrm{d}y\, m^2(y){\cal K}_{N_{\!f}}(y);
\qquad  {\cal K}_{N_{\!f}}(y)=\left\{\left[1-2CY(y)\right]y^2 \DeltaQ^2(y)\right\}',
\label{fullmass0}
\ee
where the prime denotes derivatives with respect to $y$.
The usefulness of this relation lies in the fact that, when used in 
combination with the unquenched propagators obtained following the approach of~\cite{Aguilar:2012rz}, 
it allows one to deduce how the running mass will be affected by the presence of dynamical fermions.  

\begin{figure}[!t]
\includegraphics[scale=1]{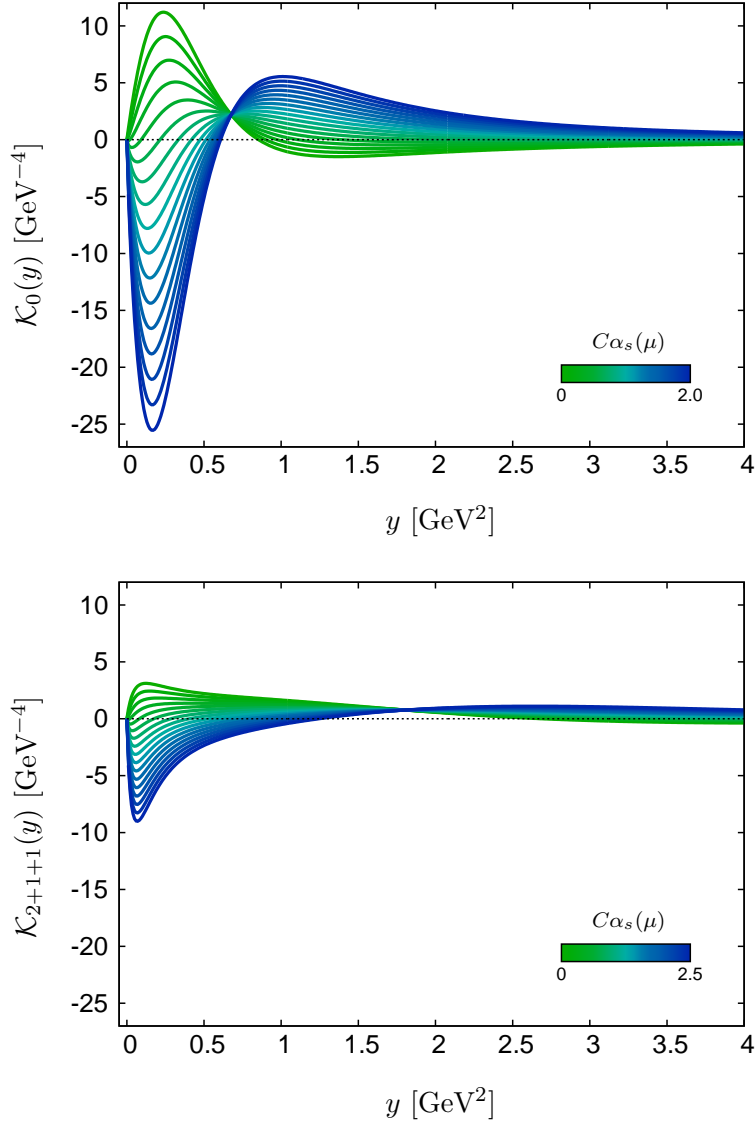}
\caption{\label{kernels} (color online) Comparison between the kernel ${\cal K}_{N_{\!f}}(y)$ of~\1eq{fullmass0} for the quenched case (top), and the unquenched case with $\nf=2+1+1$ (bottom).  The plots are drawn with the same scale to facilitate the comparison.} 
\end{figure}

To see how this works in detail, consider first the quenched case; the   kernel ${\cal K}_0$, 
constructed from the corresponding lattice propagator data, is shown on the top panel of~\fig{kernels}. 
As can be appreciated there, for increasing values of  $C$ one observes the appearance of a negative area, in the 
region of momenta $q^2<1$ GeV$^2$. 
It is precisely the contributions from this negative region that counteract the 
minus sign in front of the mass  equation~\noeq{masseq},  
thus making possible the  existence of positive-definite and monotonically decreasing solutions~\cite{Binosi:2012sj}.


Let us now repeat this exercise for the unquenched case with $\nf=2+1+1$.
On the bottom panel of~\fig{kernels}  we show the kernel ${\cal K}_{2+1+1}$; 
the $\Delta_{2+1+1}(q^2)$  used 
to construct it is obtained from ~\1eq{mastformeuc}, using the $\widehat{X}$ that corresponds to 
the aforementioned quark configuration, and setting $\lambda^2=0$.
As we will see in the next section, this particular propagator emerges precisely after the 
first step of the iterative procedure employed to solve the SDE system.
Note also that, due to the  decoupling of the heavy fermions~\cite{Aguilar:2012rz}, the 
kernel ${\cal K}_{2}$, corresponding to the case $\nf=2$, is practically identical to ${\cal K}_{2+1+1}$.

As it is evident from~\fig{kernels}, the addition of dynamical fermions 
affects the shape of the kernel significantly, as it now displays 
a much shallower negative region compared to the quenched case. 
Once this kernel is inserted into the mass equation (for the successive iterative steps), one then observes that 
in order to keep satisfying the constraint~\1eq{fullmass0} with a positive $m^2(0)$, the dynamical  mass must 
increase (and therefore the saturation point of the unquenched propagator decrease), in order to compensate for the suppressed negative region. 
This increase is counteracted in part 
by the fact that the coupling constant $\alpha_s$ also rises 
when adding quarks [at $\mu=4.3$ GeV, one has a 30\% increase 
for the $\nf=2$ case (from 0.22 to 0.285),  
and an additional 20\% increase (from 0.285 to 0.33) for the $\nf=2+1+1$ case]. 
  
This qualitative conclusion is supported by both the SDE study of~\cite{Aguilar:2012rz} as well as the lattice results of~\cite{Ayala:2012pb}, 
and will be fully confirmed through the explicit numerical computations carried out in the next subsection 
(see in particular \fig{gluonmass_fig}, where the final running masses for the different cases are presented).

\subsection{Solving the system and comparing with the lattice}

The system of SDEs that we consider is composed 
of~\3eqs{masseq}{quark-loop-full}{mastformeuc}, supplemented by the
quark  gap equation.  The initial  condition is provided  by the
quenched $SU(3)$ gluon propagator and ghost dressing function obtained
in the lattice  simulations of~\cite{Bogolubsky:2009dc}, which will be
also used to  determine the initial values of  the form factors $A(p)$
and  ${\cal M}(p)$. All  calculations will  be performed  at $\mu=4.3$.

The algorithm that we adopt consists of  the following main steps.

\n{i} Using the  quenched propagator  as an input  of the  first iterative
step, one determines the quark form factors $A(p)$ and ${\cal M}(p)$ by
solving  the quark  gap  equation;
\n{ii} $A(p)$ and ${\cal M}(p)$ are
substituted  into~\1eq{quark-loop-full}, and  the corresponding  value of
the  quark loop diagram  $\widehat{X}(q^2)$ is evaluated;
\n{iii} The preliminary form of $\DeltaQ(q^2)$  is determined from~\noeq{mastformeuc},              
employing            initially
$\lambda^2(q^2)=0$, with the  quenched mass $m^2(q^2)$ obtained
from the  solution of the  mass equation~\noeq{masseq} corresponding
to  the  quenched   lattice  propagator;  
\n{iv} The unquenched propagator 
$\DeltaQ(q^2)$ of the previous step   is substituted    into   the   mass
equation~\noeq{masseq}   in  order   to  determine   the  associated
unquenched dynamical  gluon mass $\mQ(q^2)$,  and therefore the
corresponding $\lambda^2(q^2)$; 
\n{v} At this point the  latter quantity is
inserted back  into the master  equation~\noeq{mastformeuc}, and the
loop starts again, until convergence, determined by the 
stability of the quantities involved, has been reached. 


\begin{figure}[!t]
\includegraphics[scale=1]{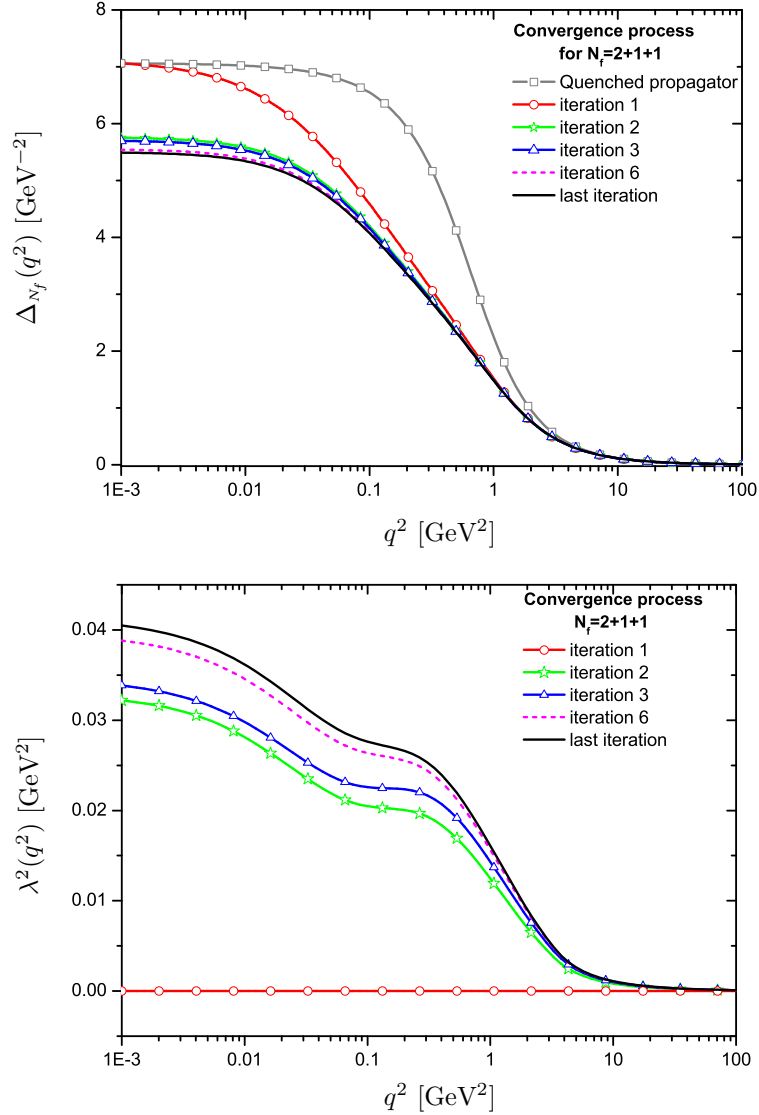}
\caption{\label{conv_fig}(color online)  Convergence of the iterative procedure for solving the SDE system in the $\nf=2+1+1$ case. 
Top: unquenched propagator $\DeltaQ(q^2)$; Bottom:  mass difference $\lambda^2(q^2)$.} 
\end{figure}



\begin{figure}[!t]
\vspace{-1cm} 
\includegraphics[scale=.85]{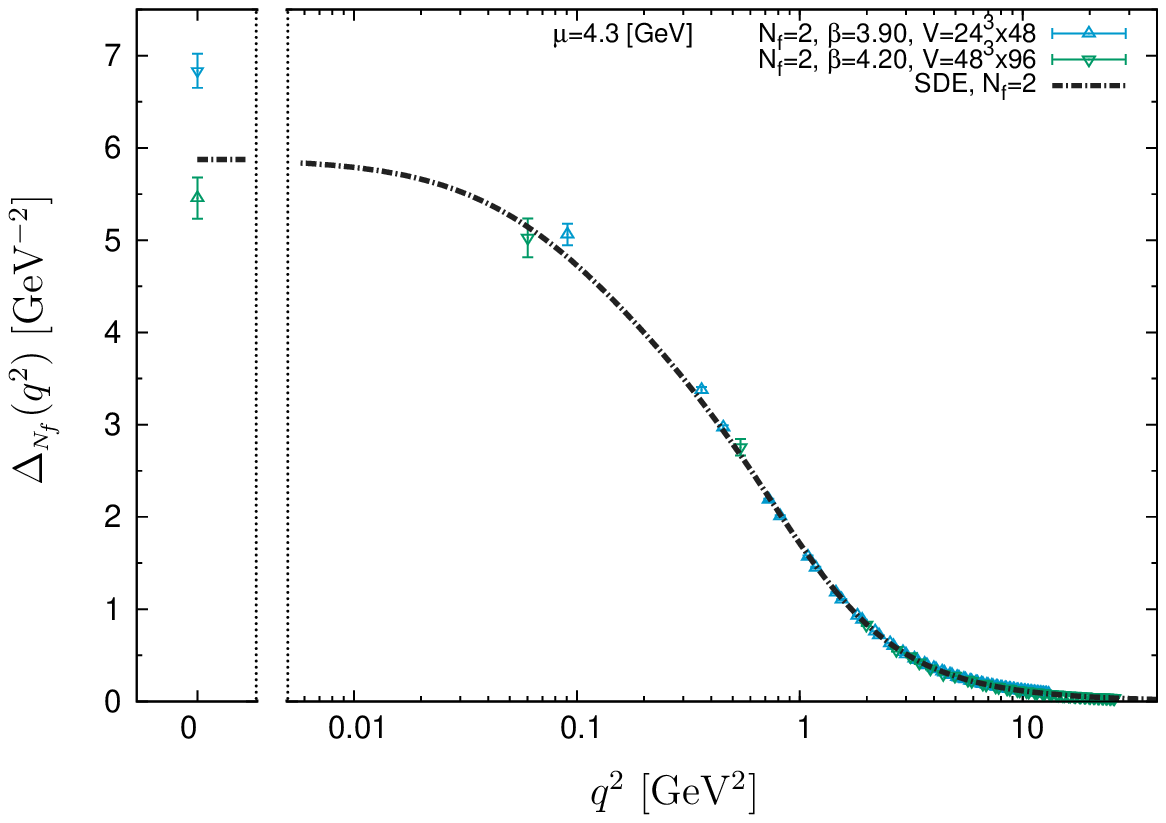}\vspace{-1cm}
\includegraphics[scale=.85]{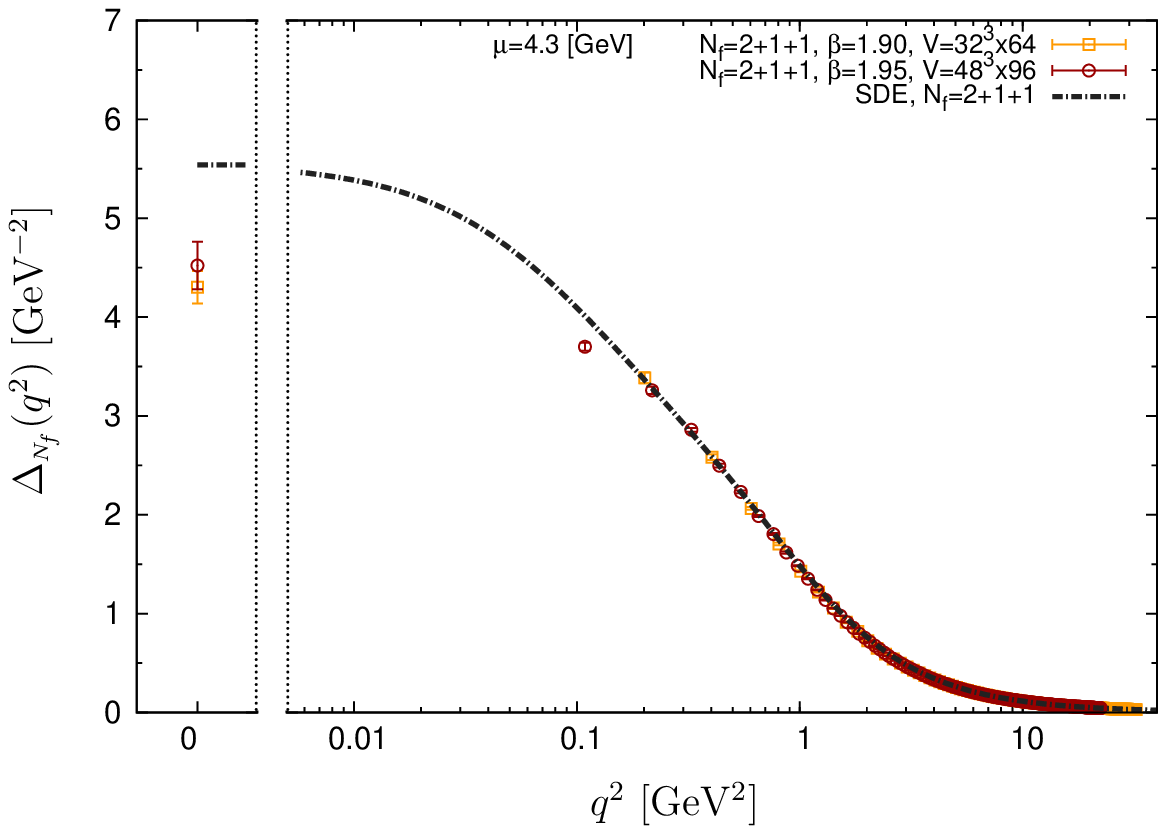}
\caption{\label{res_fig}(color online) The unquenched gluon propagators obtained from our analysis, 
for  $\nf=2$ (upper panel) and $\nf=2+1+1$ (lower panel), compared with the lattice data of~\cite{Ayala:2012pb} for the same cases.} 
\end{figure}



\begin{figure}[!t]
\includegraphics[scale=.75]{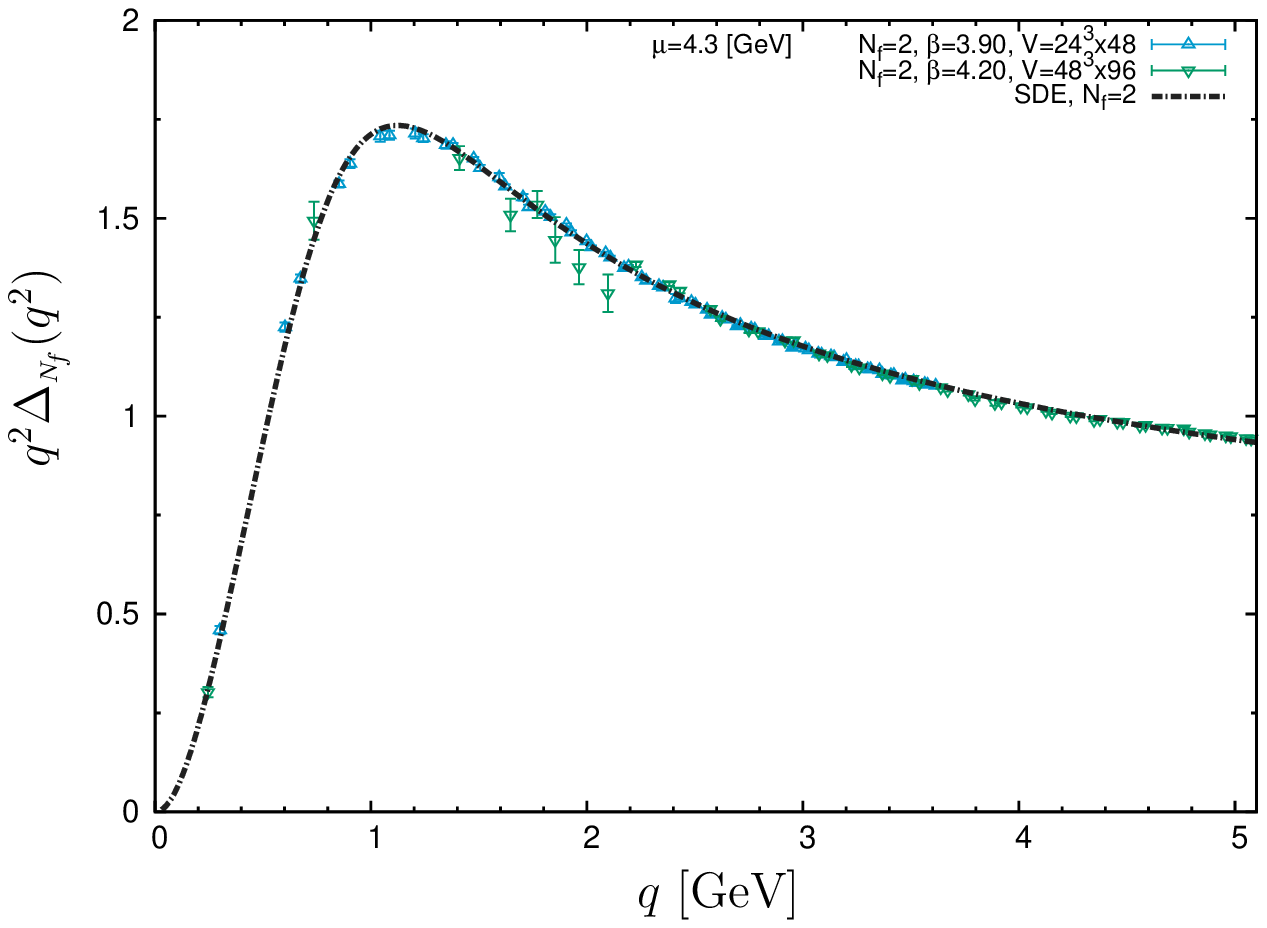}\\
\includegraphics[scale=.75]{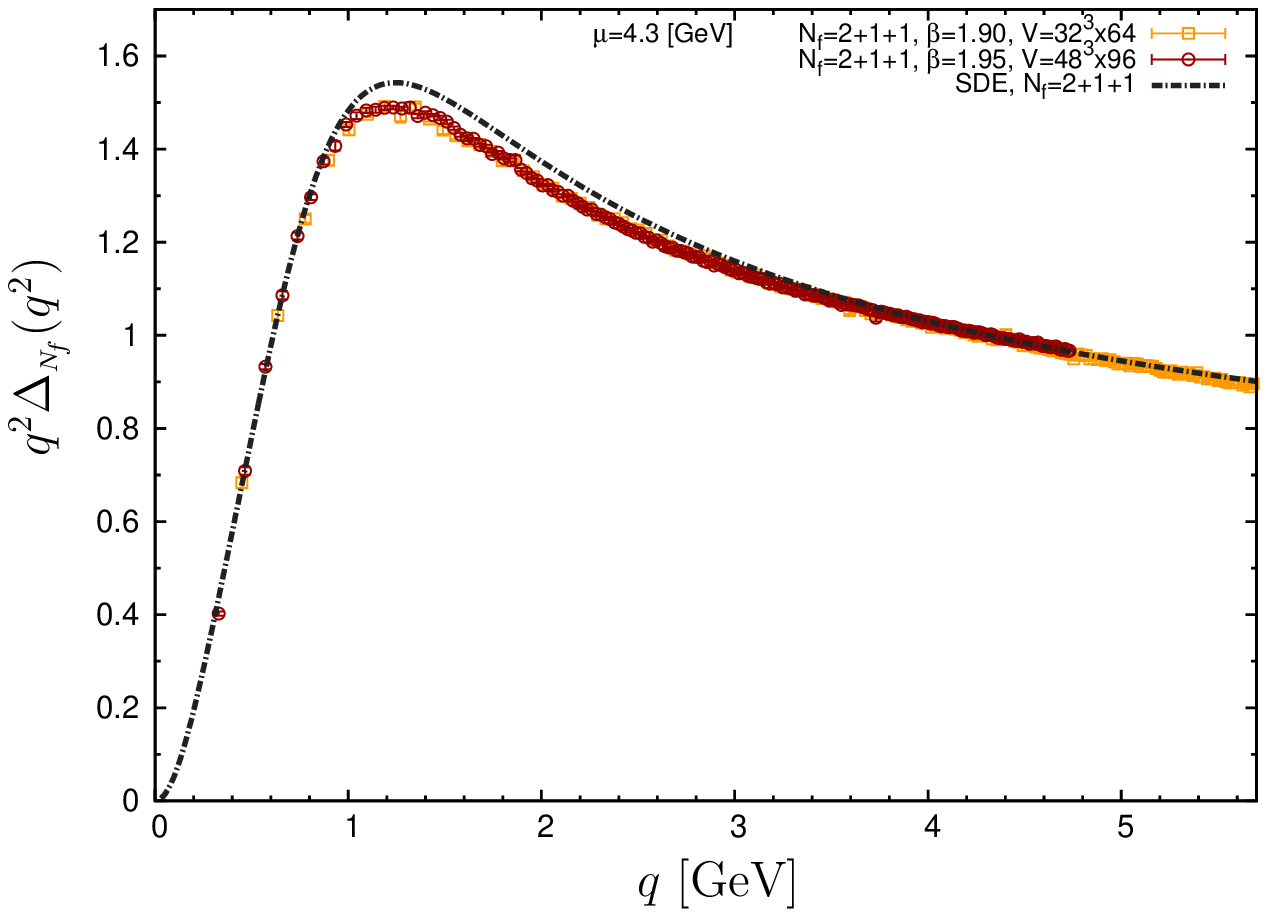}
\caption{\label{res_dr_fig}(color online)  The gluon dressing functions, $q^2\DeltaQ(q^2)$, corresponding to the propagators of the previous figure.}
\end{figure}



\begin{figure}[!t]
\includegraphics[scale=1]{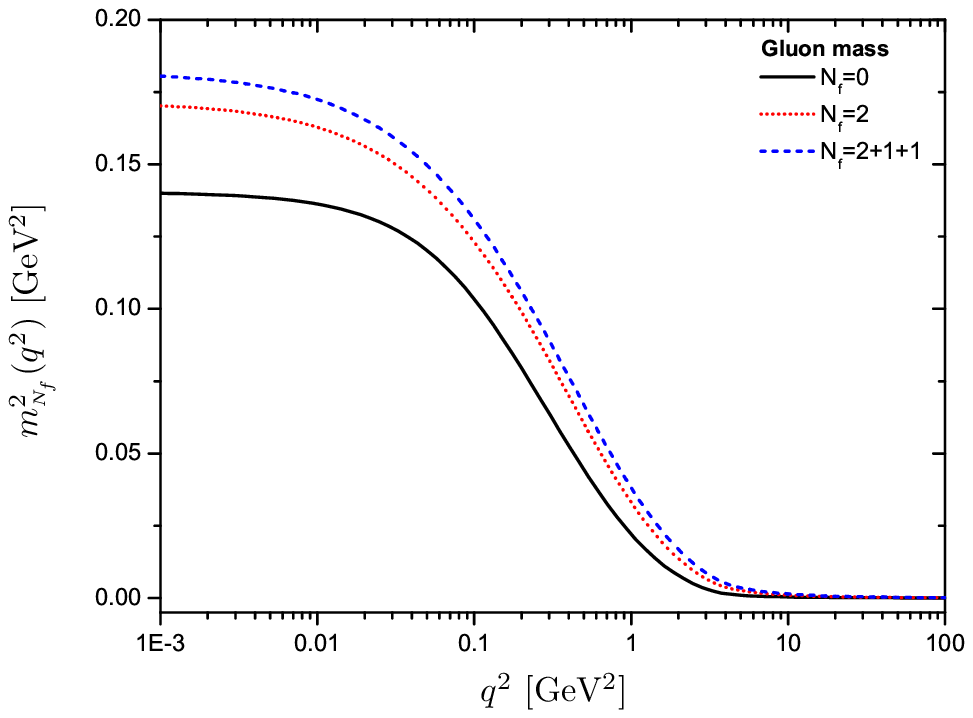} 
\caption{\label{gluonmass_fig}(color online)  Solution of the mass equation yielding the dynamically generated gluon mass for $\nf=2$ (red dotted line) and $\nf=2+1+1$ (blue dashed line). In the deep infrared one has $m_2(0)=413$ MeV, and $m_{2+1+1}(0)=425$ MeV. For comparison we also show the quenched gluon mass (black continuous line) obtained from the quenched lattice propagator, in which case $m(0)=376$ MeV.} 
\end{figure}


In~\fig{conv_fig} we show how the procedure described above converges rather rapidly to a stable solution for 
both $\DeltaQ(q^2)$ (top panel) as well as $\lambda^2(q^2)$ (bottom panel). In particular, 
observe that at the first iterative step one has $\lambda^2(q^2)=0$; even though this gives rise to 
a propagator with the same saturation point as in the quenched case (red circles curve on the top panel), 
the presence of the fermion loops alter in a rather marked way the overall shape of the function. 
As discussed in the previous section, this affects in turn the kernel of the mass equation, resulting in a $\lambda^2(q^2)>0$ 
already at the second step, which then forces the unquenched propagator to have a lower saturation point with respect to 
the quenched case (green stars curves on both panels).
Notice that, as anticipated, $\lambda^2(q^2)$ increases at each successive iteration, resulting in an unquenched 
propagator that is suppressed in the infrared with respect to the quenched case.

In~\fig{res_fig} we present the central result of this analysis. In particular,  
we plot the  propagators obtained  
when convergence of the above mentioned iteration procedure has been reached,  
and compare them with  the  corresponding  unquenched  lattice data,  recently  reported
in~\cite{Ayala:2012pb}.   We  observe   a rather good agreement
between our  theoretical predictions  and the lattice  computation for
both  values  of  $\nf$, for the available range of physical momenta. 
A notable exception to this fair coincidence between curves is the   
saturation  point  of the $\nf=2+1+1$ case; specifically, 
the value obtained from our SDE analysis is  20\% higher than that  found 
in lattice simulations. Similar conclusions can be drawn by observing the 
plot corresponding to the gluon dressing function $q^2\DeltaQ(q^2)$ shown in \fig{res_dr_fig}: 
while one has an excellent agreement in the case of two degenerate light quarks, when two heavier quarks are added the SDE solution tends to mildly 
overestimate the amplitude of the characteristic peak located in the intermediate  momentum region.

The corresponding dynamical gluon masses,  $m^2_2(q^2)$ and $m^2_{2+1+1}(q^2)$, are shown in~\fig{gluonmass_fig}; 
for comparison,  we also plot the quenched solution, $m^2(q^2)$, obtained  
from \1eq{meq} when the quenched lattice propagators of~\cite{Bogolubsky:2009dc} are used as input. 
In particular, the corresponding 
saturation points give $m_2(0)=413$ MeV and $m_{2+1+1}(0)=425$ MeV (at $\mu=4.3$ GeV), 
which should be compared with the value $m(0)=376$ MeV obtained for the quenched case.
The results captured in ~\fig{gluonmass_fig} are particularly important, because they demonstrate clearly 
that the mass generation mechanism established for pure Yang-Mills 
continues to operate in QCD-like circumstances.

Let us finally comment on some possible factors that might account for 
the aforementioned discrepancies between our predictions and the lattice data for $\nf=2+1+1$.

At first sight, one might be tempted to attribute part of the observed discrepancy to  
the presence of volume effects. Specifically,  
the propagator  plot of~\fig{res_fig}  seems to
indicate that the deviation happens in the deep IR region $q^2\lesssim
0.2$ GeV$^2$,  where the lattice data  are expected to  be affected by
moderate    volume    artifacts     (see    in    particular    Fig.~2
of~\cite{Ayala:2012pb}).  
However, a closer look at the corresponding gluon
dressing  function ( plot  of~\fig{res_dr_fig}) advocates against this possibility; 
indeed, one observes the onset of a noticeable discrepancy
as  early as $q^2\sim  2.8$ GeV$^2$, namely at a point where volume  effects are
well  under control. 
Therefore it  is reasonable  to assume  that the
observed deviation is not lattice related, 
but signals rather a mild  violation of one of the assumptions
underlying      the       derivation      of      the      unquenching
formula~\noeq{mastformeuc}.
 
In  particular, it is natural to expect that one of our  
basic operating assumption, namely  that 
the quark-loop contributions constitute a ``perturbation''  of   
the   quenched propagator, will become progressively less accurate as the number of active flavors increases. It is therefore possible that from $\nf>2$ onward we begin to perceive the onset of additional effects, not captured by~\noeq{mastformeuc}.
For instance, the identification  of the  ``pure'' gluonic diagrams of the gluon SDE  with the  gluon propagator $\Delta (q^2)$ obtained in quenched lattice simulations,  may no longer be entirely reliable, 
due to the presence of (increasingly appreciable) effects,  coming from higher-order quark subdiagrams. 

Clearly, an additional theoretical uncertainty originates from the 
approximate (perturbative) treatment of the quantity $Y(k^2)$. In particular, the parameter $C$  
may only model, to some extent, unknown contributions that display 
a logarithmic momentum dependence, 
as in \1eq{Yappr}, but cannot account for terms with a different functional form. 
Moreover, the use of an Ans\"atz for the vertex $\widehat\Gamma_{\mu}$  
entering into the definition of ${\widehat X}(q^2)$ may 
induce further error, due to the fact that its 
transverse (automatically conserved part) is in general undetermined.

\section{\label{conc}Conclusions}

In this work we addressed the basic theoretical question of whether the 
mass generation mechanism established in pure Yang-Mills
studies persists in real-world QCD.  
Specifically, by  employing a methodology  relying mainly on  the SDEs
that describe the gluon two-point sector within the PT-BFM  framework, 
we  studied in quantitative detail how the 
inclusion of dynamical quarks
affects  the generation  of  the momentum-dependent  gluon mass, in the 
Landau gauge.  
This important issue is especially relevant and timely, given 
the qualitative picture that appears to emerge from recent 
unquenched lattice simulations. 
Our results demonstrate clearly that the 
gluon  propagator  still
saturates in the infrared (therefore, a gluon mass is indeed generated), 
and that the saturation point 
is progressively suppressed, as the number of quark flavors
increases.

Several of the salient dynamical features pertinent to 
the transition from the  quenched to the
unquenched theory have been presented in  the preliminary
study of~\cite{Aguilar:2012rz}.  
The main  novelty of
the present analysis resides in the fact that, unlike~\cite{Aguilar:2012rz}, 
where  the saturation  point  of the  gluon
propagator was estimated by means of an extrapolation
procedure,  here it is determined explicitly from the 
solution of the gluon  mass equation. 
Therefore, the results obtained constitute
a genuine prediction of the formalism employed, 
emerging entirely from the inherent dynamical equations.

As has been explained in~\cite{Aguilar:2012rz} and in the present work,
the ``lowest order unquenching'' consists in 
including explicitly the contribution of the quark loop $\widehat{X}(q^2)$, keeping all other quantities unquenched. 
This is reflected clearly at the level of the master formula \1eq{mastformeuc}, where the quantity $1+G(q^2)$ 
(or, equivalently, $F^{-1}(q^2)$, by virtue of \1eq{funrelapp}) 
assumes its quenched form, obtained from \cite{Bogolubsky:2009dc}.  
Moreover, the computation of $\widehat{X}(q^2)$  [see \fig{Unq-gl-SDE}] uses as input the  
quark propagator obtained from the gap equation, which, in turn, depends on both the gluon propagator 
and the ghost dressing function; again, the quenched forms of \cite{Bogolubsky:2009dc} were employed.
Finally, the strength of the gauge coupling $g$ also depends on the number of flavors; in the present analysis 
we have used its value when $\nf=0$ (momentum-subtraction scheme~\cite{Boucaud:2008gn}).
In order to improve this analysis, and 
eventually reach a better agreement with the lattice, one possible strategy would be 
to gradually introduce quark effects into some of 
the aforementioned (quenched) ingredients.
For example, one could envisage the possibility of using unquenched instead of quenched data for the ghost dressing function 
$F(q^2)$, obtained from the lattice analysis of~\cite{Ayala:2012pb}. Given that this quantity enters both 
in the master formula and the gap equation, its overall effect may be appreciable. 
In addition, the increase 
in the value of the gauge coupling produced by the inclusion of quark flavors may modify our predictions in the 
right directions. We hope to be able to implement some of these improvements in the near future.

It is worth pointing out that the ingredients obtained from the present analysis may be used to 
determine the flavor-dependence of the QCD effective charge, ${\overline\alpha}_s(q^2)$.  
Specifically, as has been shown in~\cite{Aguilar:2009nf}, this latter quantity may be defined in 
terms of the functions $J(q^2)$ and $G(q^2)$, through the renormalization-group invariant 
combination \mbox{${\overline\alpha}^{\,-1}_s(q^2) = \alpha_s^{-1}(\mu^2) [1+G(q^2;\mu^2)]^{2} {J_m}(q^2;\mu^2)$}, 
where $\mu$ is the renormalization (subtraction) point chosen, within an appropriate renormalization scheme. 
The direct determination of $J(q^2)$ from its own dynamical equation, shown schematically in~\1eq{meq}, 
is thwarted from the fact that the main ingredients entering in it are the fully dressed three-gluon 
vertex  $BQ^2$ (studied in~\cite{Binosi:2011wi}) and the (largely unexplored) four-gluon vertex $BQ^3$, 
for arbitrary values of their momenta. 
Instead, one may employ  \1eq{massiveprop} to  compute $J(q^2)$ indirectly,  
using a combined approach based on the  knowledge of the full propagator from the lattice, 
and the  corresponding gluon mass from solving \1eq{masseq}.  
(For the possibility of the direct extraction of 
the quantity \mbox{${\widehat J}_m (q^2;\mu^2)= [1+G(q^2;\mu^2)]^{2} {J_m}(q^2;\mu^2)$} 
from specialized lattice simulations, see~\cite{Binosi:2013rba}). Work in this direction is already in progress.

Finally, it would clearly be important to study 
the infrared dynamics of the gluon propagator, quenched and unquenched, 
away from the Landau gauge. 
The SDEs formulated within the PT-BFM framework 
provide a natural starting point for such an investigation in the 
continuum. In fact, it would be particularly relevant to 
determine whether the detailed mass generation mechanism
established in the Landau gauge persists for other values of
the gauge-fixing parameter, such as, for example, the 
Feynman gauge, which is known to be of central importance for the PT.
 To be sure, further reliable input from the lattice would be invaluable for 
accomplishing such a demanding task, given that, 
even though the formal procedure for 
implementing covariant $R_\xi$ gauges on the lattice has been already 
reported~\cite{Cucchieri:2009kk},  the available simulations 
are restricted to  modest    size   volumes,   and 
$\xi\ll1$ only~\cite{Cucchieri:2011pp}. 
We hope to report progress in some of these directions in the near future. 

\acknowledgments 

The research of J.~P. is supported by the Spanish MEYC under 
grant FPA2011-23596. The work of  A.~C.~A  is supported by the 
National Council for Scientific and Technological Development - CNPq
under the grant 306537/2012-5 and project 473260/2012-3,
and by S\~ao Paulo Research Foundation - FAPESP through the project 2012/15643-1.


\end{document}